\documentclass[10pt,twocolumn,letterpaper]{article}

\usepackage[pagenumbers]{cvpr} 

\usepackage{graphicx}
\usepackage{amsmath}
\usepackage{amssymb}
\usepackage{booktabs}
\usepackage{listings}

\usepackage[pagebackref,breaklinks,colorlinks]{hyperref}

\usepackage[capitalize]{cleveref}
\crefname{section}{Sec.}{Secs.}
\Crefname{section}{Section}{Sections}
\Crefname{table}{Table}{Tables}
\crefname{table}{Tab.}{Tabs.}

\def\cvprPaperID{*****} 
\def\confName{CVPR}
\def\confYear{2022}

\begin{document}

\title{HoloLens 2 Sensor Streaming}

\author{Juan Carlos Dibene\\
Stevens Institute of Technology\\
{\tt\small }
\and
Enrique Dunn\\
Stevens Institute of Technology\\
{\tt\small }
}
\maketitle

\begin{abstract}
We present a HoloLens 2 server application for streaming device data via TCP in real time.
The server can stream data from the four grayscale cameras, depth sensor, IMU, front RGB camera, microphone, head tracking, eye tracking, and hand tracking.
Each sent data frame has a timestamp and, optionally, the instantaneous pose of the device in 3D space.
The server allows downloading device calibration data, such as camera intrinsics, and can be integrated into Unity projects as a plugin, with support for basic upstream capabilities.
To achieve real time video streaming at full frame rate, we leverage the video encoding capabilities of the HoloLens 2.
Finally, we present a Python library for receiving and decoding the data, which includes utilities that facilitate passing the data to other libraries. The source code, Python demos, and precompiled binaries are available at \url{https://github.com/jdibenes/hl2ss}.
\end{abstract}


\section{Introduction}

In this report, we present a real time system for streaming HoloLens 2 sensor data over TCP.
The system consists of a server application that runs on the HoloLens 2 and a Python client library that receives and decodes the data.
The client library works on Windows, Linux, and OS X systems.
The server can also be integrated into Unity projects as a plugin and has support for receiving messages from the client.
This allows leveraging the compute power of the client and the powerful capabilities of the Unity Engine on HoloLens 2.

The purpose of this system is to enable real time online experiments with HoloLens 2 data on other systems, with potentially more compute capabilities and third-party library support.
The Windows Device Portal \cite{uwp_device_portal} can stream data from the HoloLens 2 front RGB camera and the microphone, but does not provide access to the four side-view grayscale cameras, the depth sensor, and the IMU.
The Research Mode API \cite{hl2_rm} provides access to these sensors, but the given tools only allow recording the data to files and then downloading them from the HoloLens 2.

\begin{figure}[ht!]
    \centering
    
    \begin{minipage}{\linewidth}
        \centering
        \includegraphics[width=\linewidth]{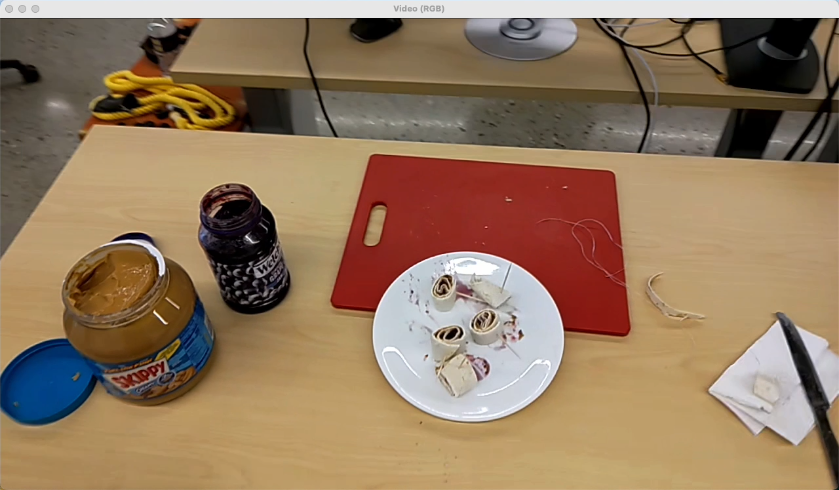} 
    \end{minipage}
    
    \begin{minipage}{0.48\linewidth}
        \centering
        \includegraphics[width=\linewidth]{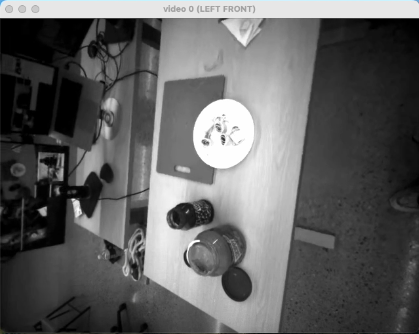} 
    \end{minipage} \hfill
    \begin{minipage}{0.48\linewidth}
        \centering
        \includegraphics[width=\linewidth]{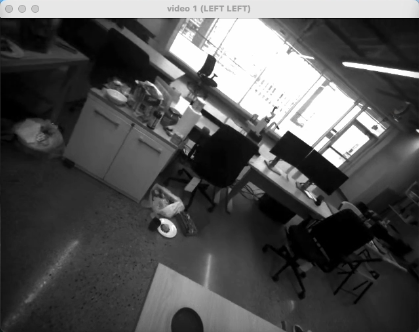} 
    \end{minipage}

    \begin{minipage}{0.48\linewidth}
        \centering
        \includegraphics[width=\linewidth]{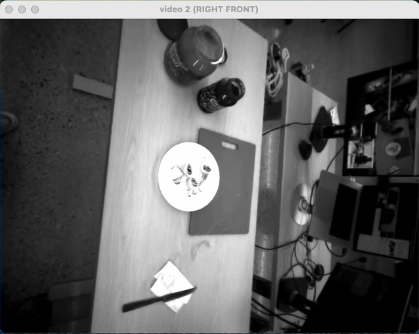} 
    \end{minipage} \hfill
    \begin{minipage}{0.48\linewidth}
        \centering
        \includegraphics[width=\linewidth]{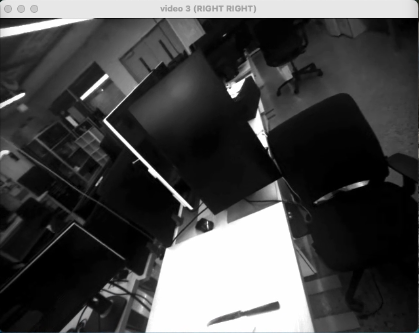} 
    \end{minipage}

    \begin{minipage}{0.48\linewidth}
        \centering
        \includegraphics[width=\linewidth]{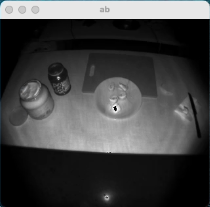} 
    \end{minipage} \hfill
    \begin{minipage}{0.48\linewidth}
        \centering
        \includegraphics[width=\linewidth]{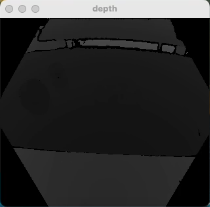} 
    \end{minipage}

   \caption{Our HoloLens 2 Sensor Streaming system allows to stream HoloLens 2 sensor data over Wi-Fi, enabling real time, online experiments on Windows, Linux, and OS X systems.}
   \label{fig:teaser}
\end{figure}

Our system fills this gap by providing access to all of these sensors, and to head pose, eye tracking, and hand tracking data over Wi-Fi. \autoref{fig:teaser} shows simultaneous capture from all of the HoloLens 2 cameras using our system.
Data is transferred in real time at full framerate with low latency.
For 1080p 30 FPS video from the RGB camera, we measured a latency of approximately 270 ms in our setup.

The rest of this report is organized as follows. Section \ref{sec:server} describes the details of our HoloLens 2 server application.
Section \ref{sec:unity} describes the features of the Unity plugin.
Section \ref{sec:pylib} introduces our Python library for data reception.

\section{HoloLens 2 Server Application}
\label{sec:server}

Our HoloLens 2 server is a C++ Universal Windows Platform (UWP) application that exposes the following data streams at specific TCP ports as shown in \autoref{tab:tcpports}:
\begin{itemize}
    \item RM Visible Light Cameras (VLC) - Four grayscale cameras operating at 640x480 @ 30 FPS. H264 or HEVC encoded.
    \item RM Depth Long-throw - 16-bit depth (in millimeters) and 16-bit active brightness (AB) image of 320x288 @ 1-5 FPS. Encoded as a single 32-bit PNG. RM Depth AHaT is not supported.
    \item RM IMU - Accelerometer ($m/s^2$), gyroscope ($deg/s$), and magnetometer.
    \item Photo/Video RGB camera (PV) - 1920x1080 @ 30 FPS with configurable resolution, framerate, exposure, focus, and white balance. H264 or HEVC encoded.
    \item Microphone - 2 channels and 48000 Hz sample rate. Encoded as AAC ADTS.
    \item Spatial input - Head pose, gaze ray (origin and direction), and hand tracking (2 hands with 26 joints each).
\end{itemize}

Data from the grayscale cameras, depth sensor, and IMU is acquired using the Research Mode (RM) API \cite{hl2_rm}.
PV camera video is obtained using the MediaCapture class \cite{uwp_mediacapture}.
Microphone audio is captured using WASAPI \cite{uwp_wasapi}.
Head pose and eye tracking data is acquired using the SpatialPointerPose class \cite{uwp_spatialpointer}.
Hand tracking data is obtained using the SpatialInteractionManager class \cite{uwp_spatialinteraction}.

\begin{table}
  \centering
  \begin{tabular}{ll}
    \toprule
    Stream & TCP Port \\
    \midrule
    RM VLC Left-front & 3800  \\
    RM VLC Left-left & 3801 \\
    RM VLC Right-front & 3802 \\
    RM VLC Right-right & 3803 \\
    RM Depth Long-throw & 3805 \\
    RM IMU Accelerometer & 3806 \\
    RM IMU Gyroscope & 3807 \\
    RM IMU Magnetometer & 3808 \\
    PV & 3810 \\
    Microphone & 3811 \\
    Spatial Input & 3812 \\
    \bottomrule
  \end{tabular}
  \caption{TCP Ports for each data stream exposed by the HoloLens 2 server application.}
  \label{tab:tcpports}
\end{table}

\subsection{Data Compression}

To reduce network bandwidth requirements and achieve real time data streaming, audio and video are compressed using the Microsoft Media Foundation SDK \cite{uwp_mediafoundation}.
For video, the client can select one of four video encoding profiles and directly set the bitrate.
For audio, the client can select one of four bitrate presets.
All of the available encoding profiles are shown in \autoref{tab:encoding} and all of them are lossy.
For depth, the invalid pixels in the depth image, as indicated by the 8-bit 320x288 sigma channel, are set to zero.
Then, the depth and AB images are interleaved using NEON \cite{neon} and encoded as PNG (lossless) using the BitmapEncoder class \cite{uwp_bitmapencoder}.
This is 1) to avoid sending the sigma channel, and 2) interleaving the images gives better compression results than concatenating them, although this is scene-dependent.
The average bandwidth for each stream is shown in \autoref{tab:bauds}.

\begin{table}
  \centering
  \begin{tabular}{lll}
    \toprule
    ID & Video encoding profiles & Audio encoding presets \\
    \midrule
    0 & H264 Baseline    & AAC, 12000 bytes/s \\
    1 & H264 Main        & AAC, 16000 bytes/s \\ 
    2 & H264 High        & AAC, 20000 bytes/s \\
    3 & H265 Main (HEVC) & AAC, 24000 bytes/s \\
    \bottomrule
  \end{tabular}
  \caption{Available audio and video encoding profiles. All of these encoding formats are lossy.}
  \label{tab:encoding}
\end{table}

\begin{table}
  \centering
  \begin{tabular}{ll}
    \toprule
    Stream & Bandwidth \\
    \midrule
    Single RM VLC & 1 Mbit/s (default) \\
    RM Depth Long-throw & 6.5 Mbit/s \\ 
    RM IMU Accelerometer & 250 Kbit/s \\
    RM IMU Gyroscope & 1.5 Mbit/s \\
    RM IMU Magnetometer & 12 Kbit/s \\
    PV & 5 Mbit/s (default) \\
    Microphone & 192 Kbit/s (default) \\
    Spatial Input & 1 Mbit/s \\
    \midrule
    Total & 19 Mbit/s \\
    \bottomrule
  \end{tabular}
  \caption{Average bandwidth for each stream. The bitrate of the audio and video streams is configurable.}
  \label{tab:bauds}
\end{table}

\subsection{Data Transfer}

Data is transferred over TCP using Windows Sockets \cite{uwp_winsock}.
All data is transferred in little-endian format and matrices in row-major order.
The structure of a data frame, shown in \autoref{tab:packet}, is the same for all streams.
The timestamps are expressed in hundreds of nanoseconds and are in the same domain (QPC \cite{uwp_qpc}), so they can be used to correlate the data of different streams.
The device pose field is optional and its presence is controlled by the client.
The pose is expressed as a 4x4 float matrix.
If the pose is valid, the last element of the matrix is 1.
Otherwise, if the HoloLens 2 tracking is lost for that frame, the last element is 0.

Each stream runs on its own thread, so multiple streams can be active simultaneously.
However, there is no provision for the stream-client synchronization required for data frame unpacking other than the start of the stream.
Therefore, only one client per stream is allowed.
The streams support up to three different operating modes, as shown in \autoref{tab:modes}.
The client sets the stream operating mode when opening the stream and persists until the stream is closed.
In Mode 0, only the device data is transferred.
In Mode 1, the corresponding device pose for each frame is included.
Mode 2 is used by the client to download calibration data, which is device dependent, and the stream is automatically closed at the end of the transfer.
The operating modes supported by each stream are shown in \autoref{tab:supportedmodes}.

\begin{table}
  \centering
  \begin{tabular}{lll}
    \toprule
    Field & Type & Length in bytes \\
    \midrule
    Timestamp & u64 & 8 \\
    Size of payload & u32 & 4 \\
    Payload & bytes & Size of payload \\
    Pose (optional) & 4x4 float & 64 \\
    \bottomrule
  \end{tabular}
  \caption{Common data frame structure. The pose field is optional, and if not included, its length is zero. Data frames are unpacked from the stream by a simple FSM.}
  \label{tab:packet}
\end{table}

\begin{table}
  \centering
  \begin{tabular}{ll}
    \toprule
    Mode & Description \\
    \midrule
    0 & Continuous transfer of device data  \\
    1 & Continuous transfer of device data and pose  \\
    2 & Single transfer of calibration data \\
    \bottomrule
  \end{tabular}
  \caption{Stream operating modes. The client configures the operating mode when opening the stream.}
  \label{tab:modes}
\end{table}

\begin{table}
  \centering
  \begin{tabular}{ll}
    \toprule
    Stream & Supported Modes \\
    \midrule
    RM VLC & 0, 1, 2  \\
    RM Depth & 0, 1, 2 \\
    RM IMU Accelerometer & 0, 1, 2 \\
    RM IMU Gyroscope & 0, 1, 2 \\
    RM IMU Magnetometer & 0, 1 \\
    PV & 0, 1, 2 \\
    Microphone & 0 \\
    Spatial Input & 0 \\
    \bottomrule
  \end{tabular}
  \caption{Stream operating modes supported by each stream. Not all streams support all operating modes.}
  \label{tab:supportedmodes}
\end{table}

\subsection{Mode 0 and Mode 1 Transfers}
\label{sec:mode0mode1}

For all but the Spatial Input stream, the server waits for the client to send configuration data and does not begin streaming until it is received.
Configuration data is sent over the stream's TCP port in little-endian format.
The configuration parameters and their order for each stream are:
\begin{itemize}
    \item RM VLC: operating mode (u8), width (u16), height (u16), framerate (u8), video encoding profile (u8), bitrate (u32). Width, height, and framerate must be 640, 480, and 30, respectively.
    \item RM Depth: operating mode (u8).
    \item RM IMU: operating mode (u8).
    \item PV: operating mode (u8), width (u16), height (u16), framerate (u8), video encoding profile (u8), bitrate (u32). Width, height, and framerate must be one of the configurations belonging to the VideoConferencing profile shown in the locatable camera overview \cite{uwp_pv_modes}.
    \item Microphone: audio encoding preset (u8).
\end{itemize}

Data streaming begins immediately after the server finalizes processing the configuration data.
For the RM VLC and PV streams, the payload consists of encoded video frames.
For the RM Depth stream, the payload is the PNG containing both the depth (in the first 2 channels) and AB (in the last 2 channels) images.
For the RM IMU streams, the payload contains batches of samples.
The structure of each sample is shown in \autoref{tab:imusamples}.
RM IMU Accelerometer, Gyroscope, and Magnetometer frames contain 93, 315, and 11 samples, respectively.
For the Microphone stream, the payload consists of encoded audio samples.

The payload structure for the Spatial Input stream is shown in \autoref{tab:spatialinputsample}.
The set bits of the valid field indicate whether the data of the Head Pose (bit 0), Eye Ray (bit 1), Left hand (bit 2), and Right Hand (bit 3) fields are valid.
The structure of the Head Pose and Eye Ray fields are shown in \autoref{tab:headposesample} and \autoref{tab:eyeraysample}, respectively.
For the Head Pose, up corresponds to $+Y$ and forward corresponds to $-Z$.
The Left Hand and Right Hand fields contain the pose of the 26 joints of each hand.
See the HandJointKind Enum documentation \cite{uwp_hand_joint_kind} for a list of hand joints.
The structure of each hand joint pose is shown in \autoref{tab:handjointsample}.
Orientation is expressed as a quaternion.
For more details, see the documentation for the JointPose struct \cite{uwp_joint_pose}.

For Mode 1 streams, the pose of the device is included in each data frame.
For RM streams, the pose corresponds to a device-defined coordinate
frame defined as the {\em rigNode}.
See the HoloLens 2 Research Mode report \cite{hl2_rm} for details.
The extrinsics that express the transform of each sensor to the {\em rigNode} can be obtained from a Mode 2 transfer.
The poses are acquired using the SpatialLocator class \cite{uwp_spatial_locator}.
For the PV stream, the pose corresponds to the camera and comes bundled with the acquired video frames.

\begin{table}
  \centering
  \begin{tabular}{lll}
    \toprule
    Field & Type & Length in bytes \\
    \midrule
    Sensor timestamp in $ns$ & u64 & 8  \\
    Data frame timestamp & u64 & 8 \\
    X-axis measurement & float & 4  \\
    Y-axis measurement & float & 4 \\
    Z-axis measurement & float & 4 \\
    \bottomrule
  \end{tabular}
  \caption{RM IMU sample structure. The data frame timestamp is the same for all of the samples in the batch.}
  \label{tab:imusamples}
\end{table}

\begin{table}
  \centering
  \begin{tabular}{lll}
    \toprule
    Field & Type & Length in bytes \\
    \midrule
    Valid & u8 & 1 \\
    Head Pose & bytes & 36 \\
    Eye Ray & bytes & 24 \\
    Left Hand & bytes & 26x36 \\
    Right Hand & bytes & 26x36 \\
    \bottomrule
  \end{tabular}
  \caption{Spatial Input payload structure. The first 4 bits of the valid field indicate the validity of the Head Pose, Eye Ray, Left Hand, and Right Hand fields.}
  \label{tab:spatialinputsample}
\end{table}

\begin{table}
  \centering
  \begin{tabular}{lll}
    \toprule
    Field & Type & Length in bytes \\
    \midrule
    Position & 1x3 float & 12 \\
    Forward direction & 1x3 float & 12 \\
    Up direction & 1x3 float & 12 \\    
    \bottomrule
  \end{tabular}
  \caption{Head Pose structure. Vector element order is X, Y, and Z. Right direction can be obtained as cross(up, -forward).}
  \label{tab:headposesample}
\end{table}

\begin{table}
  \centering
  \begin{tabular}{lll}
    \toprule
    Field & Type & Length in bytes \\
    \midrule
    Position & 1x3 float & 12 \\
    Direction & 1x3 float & 12 \\
    \bottomrule
  \end{tabular}
  \caption{Eye Ray structure. Vector element order is X, Y, and Z. Ray length is not provided.}
  \label{tab:eyeraysample}
\end{table}

\begin{table}
  \centering
  \begin{tabular}{lll}
    \toprule
    Field & Type & Length in bytes \\
    \midrule
    Orientation & 1x4 float & 16 \\
    Position & 1x3 float & 12 \\
    Radius & float & 4 \\
    Accuracy & u32 & 4 \\
    \bottomrule
  \end{tabular}
  \caption{Hand Joint structure. Quaternion element order is X, Y, Z, and W. Vector element order is X, Y, and Z.}
  \label{tab:handjointsample}
\end{table}

\subsection{Mode 2 Transfers}

The calibration data for each sensor can be obtained from a single Mode 2 transfer.
Like Mode 0 and Mode 1 transfers, the client must send the configuration data first.
For RM streams, only the operating mode byte must be sent.
For the PV stream, the whole configuration string, as defined in \autoref{sec:mode0mode1}, must be sent.
The server automatically closes the connection after all the calibration data has been transferred.

The structure of the calibration data for the RM VLC, RM Depth Long-throw, and RM IMU streams is shown in \autoref{tab:mode2vlc}, \autoref{tab:mode2depth}, and \autoref{tab:mode2imu}, respectively.
The uv2xy LUT converts image coordinates to normalized coordinates on the camera unit plane.
First channel corresponds to $x$, second channel to $y$.
The extrinsics matrix for the sensor is the transform to the {\em rigNode}.
The scale value is used to convert depth units to meters.
This data is obtained using the Research Mode API \cite{hl2_rm}, and more details can be found in its documentation.
The structure of the calibration data for the PV stream is shown in \autoref{tab:mode2pv}.
This data is embedded in every video frame.
However, since the server only exposes this data through Mode 2 transfers, the autofocus function of the PV camera must be disabled so the downloaded calibration data remains valid.

\begin{table}
  \centering
  \begin{tabular}{lll}
    \toprule
    Field & Type & Length in bytes \\
    \midrule
    LUT uv2xy & 480x640x2 float & 2457600 \\
    Extrinsics & 4x4 float & 64 \\
    \bottomrule
  \end{tabular}
  \caption{RM VLC calibration data.}
  \label{tab:mode2vlc}
\end{table}

\begin{table}
  \centering
  \begin{tabular}{lll}
    \toprule
    Field & Type & Length in bytes \\
    \midrule
    LUT uv2xy & 288x320x2 float & 737280 \\
    Extrinsics & 4x4 float & 64 \\
    Scale & float & 4 \\
    \bottomrule
  \end{tabular}
  \caption{RM Depth Long-throw calibration data.}
  \label{tab:mode2depth}
\end{table}

\begin{table}
  \centering
  \begin{tabular}{lll}
    \toprule
    Field & Type & Length in bytes \\
    \midrule
    Extrinsics & 4x4 float & 64 \\
    \bottomrule
  \end{tabular}
  \caption{RM IMU calibration data. Not available for RM IMU Magnetometer.}
  \label{tab:mode2imu}
\end{table}

\begin{table}
  \centering
  \begin{tabular}{lll}
    \toprule
    Field & Type & Length in bytes \\
    \midrule
    Focal length & 1x2 float & 8 \\
    Principal point & 1x2 float & 8 \\
    Radial distortion & 1x3 float & 12 \\
    Tangential distortion & 1x2 float & 8 \\
    Projection & 4x4 float & 64 \\
    \bottomrule
  \end{tabular}
  \caption{PV calibration data.}
  \label{tab:mode2pv}
\end{table}

\subsection{Remote Configuration Port}

The server exposes an interface at TCP port 3809 that the client can use to send secondary configuration data and query information.
All data is in little-endian format.
The available commands and their parameters are as follows:
\begin{itemize}
    \item Set PV display marker state: 0 (u8), enable (u8). Controls the display marker used to show to the user the lower boundary of the field of view of the PV camera. This marker is used to help the user keep objects of interest within PV camera video frames.
    \item Set PV camera focus: 1 (u8), focus mode (u32), autofocus range (u32), manual focus distance (u32), focus value (u32), driver fallback (u32). See the FocusControl class \cite{uwp_focus_control} for details.
    \item Set PV camera video temporal denoising: 2 (u8), mode (u32). See the VideoTemporalDenoisingControl class \cite{uwp_temporal_denoising} for details.
    \item Set PV camera white balance preset: 3 (u8), preset (u32). See the WhiteBalanceControl class \cite{uwp_white_balance}.
    \item Set PV camera white balance value: 4 (u8), value (u32). See the WhiteBalanceControl class \cite{uwp_white_balance}.
    \item Set PV camera exposure: 5 (u8), mode (u32), value (32). See the ExposureControl class \cite{uwp_exposure} for details.
    \item Get server version: 6 (u8). Returns the server version as 1x4 vector of u16.
\end{itemize}

The PV camera can be configured even if the PV stream is transferring data to the client.
For the white balance value and exposure commands, the values must be divided by 25 and 10, respectively.
This makes the step size equal to 1.
The server undoes this transformation so the values are in the expected range.
Like the data streams, only one connection at a time is allowed.

\section{Unity Integration}
\label{sec:unity}

Our HoloLens 2 server can be integrated into Unity projects as a plugin.
All the streams are supported.
However, Spatial Input support requires that the plugin is initialized from the UI thread.
The PV display marker functionality is not supported.
Unlike the standalone server, the plugin has basic upstream capabilities.

\subsection{Unity IPC Port}

The plugin exposes an additional interface at TCP port 3816 that allows the client to send messages to a Unity application.
The message structure is shown in \autoref{tab:unityipc}.
The interface is message-agnostic, so the meanings of the commands and their parameters are defined by the user.
This allows the user to add support for new commands to their Unity application without modifying the plugin.
However, command \$FFFFFFFF is reserved and should not be used for new commands, as it is used to notify the Unity application that the client has disconnected from the IPC port.
The plugin also supports sending responses to the client in the form of 4-byte integers.
All data is in little-endian format.
Only one connection at a time is allowed.

\begin{table}
  \centering
  \begin{tabular}{lll}
    \toprule
    Field & Type & Length in bytes \\
    \midrule
    Command ID & u32 & 4 \\
    Size of parameters & u32 & 4 \\
    Parameters & bytes & Size of parameters \\
    \bottomrule
  \end{tabular}
  \caption{Unity plugin IPC message structure. Command ID \$FFFFFFFF is reserved and should not be used by the client.}
  \label{tab:unityipc}
\end{table}

\subsection{Remote Unity Scene}

As an example, we created a Unity project with the Mixed Reality Toolkit (MRTK) \cite{uwp_mrtk} and implemented a set of commands to allow the client to create basic Unity objects, which the user can see in augmented reality (AR). The commands and their parameters are:
\begin{itemize}
    \item Create primitive (0): type (u32). Creates a Unity primitive in the Unity scene. See \autoref{tab:unityprimitives}. Returns key (u32), which can be used to modify the properties of the primitive.
    \item Set active (1): key (u32), state (u32). Activates or deactivates the object associated with key.
    \item Set world transform (2): key (u32), position (1x3 float), rotation (1x4 float), scale (1x3 float). Sets the world transform of the object associated with key. Rotation is a quaternion.
    \item Set color (4): key (u32), rgba (1x4 float). Sets the color of the primitive associated with key. The elements of rgba are in $[0, 1]$, and semi-transparency is supported.
    \item Set texture (5): key (u32), texture (JPG or PNG file). Sets the texture of the primitive associated with key.
    \item Create text (6). Creates a TextMeshPro object. Returns key (u32), which can be used to modify the properties of the object.
    \item Set text (7): key (u32), font size (float), rgba (1x4 float), string (utf-8). Sets the text, font size, and color of the TextMeshPro object associated with key.
    \item Remove (16): key (u32). Destroys the object associated with key.
    \item Remove all (17). Destroys all objects created using the plugin.
    \item Begin display list (18). Hint for the Unity application to process the next commands in the same pass.
    \item End display list (19).
    \item Set target mode (20): mode (u32). Changes the behavior of commands that modify the properties of objects. If mode is 0, property changes apply to the object associated with key. If mode is 1, the key parameter is ignored and property changes apply to the last object created. This is to allow the client to create objects and set their properties immediately without having to wait for the server to return the key.
\end{itemize}

All of the commands return a 4-byte value. For the commands that do not return a key, the value 0 denotes failure, and 1 denotes success. 

\begin{table}
  \centering
  \begin{tabular}{ll}
    \toprule
    Type & Primitive \\
    \midrule
    0 & Sphere \\
    1 & Capsule \\
    2 & Cylinder \\
    3 & Cube \\
    4 & Plane \\
    5 & Quad \\
    \bottomrule
  \end{tabular}
  \caption{Available Unity primitives.}
  \label{tab:unityprimitives}
\end{table}

\section{Python Library}
\label{sec:pylib}

We offer a Python library that facilitates connecting to HoloLens 2 server, sending configuration data and commands, receiving data and decoding it, and preprocessing the data for use with other libraries.
For H264 and AAC decoding we use the PyAV \cite{pyc_pyav} library, and PNG decoding is performed using OpenCV \cite{pyc_opencv}.

The library abstracts most of the communication details and presents a simple interface to the user. Here is an example of acquiring 1080p 30 FPS video from the PV camera and displaying it in a window in real time.
\begin{verbatim}
import cv2
import hl2ss
import hl2ss_utilities

# HoloLens 2 IP address
host = '192.168.1.15'

# Stream settings
port = hl2ss.StreamPort.PERSONAL_VIDEO
chunk = hl2ss.ChunkSize.PERSONAL_VIDEO
mode = hl2ss.StreamMode.MODE_1
width = 1920
height = 1080
framerate = 30

# Video encoding settings
profile = hl2ss.VideoProfile.H265_MAIN
bitrate = 5*1024*1024

# Video decoding settings
output_format = 'bgr24'

client = hl2ss_utilities.rx_decoded_pv(
             host, 
             port,
             chunk,
             mode,
             width,
             height,
             framerate,
             profile,
             bitrate,
             output_format)

client.open()

while (enable):
    # Get next video frame, blocks
    # until the next frame is received
    data = client.get_next_packet()

    # data fields are:
    # data.timestamp
    # data.payload
    # data.pose (None for Mode 0)

    # Display frame using OpenCV
    # Decoded payload is a 1080x1920x3
    # NumPy array of u8
    cv2.imshow('Video', data.payload)
    cv2.waitKey(1)
    
client.close()
\end{verbatim}

The procedure to acquire data from other sensors is very similar: 1) create a hl2ss rx  or hl2ss\_utilities rx\_decoded (for the encoded streams) object with the desired configuration, 2) call the open() method to connect to the HoloLens 2 server, 3) continuously call the get\_next\_packet() method to receive data frames, and 4) call the close() method to close the connection.
The get\_next\_packet() method must be called as soon and as often as possible to avoid dropping frames.
Note that when using a hl2ss rx object with an encoded stream, the payload contains encoded frames. 

One important limitation is that the get\_next\_packet() method blocks until the next frame is received, which complicates working with multiple streams.
To overcome this, we provide a library extension based on Python's multiprocessing library.

\subsection{Multiprocessing Extension}

We extend our library using Python's multiprocessing capabilities to facilitate working with multiple streams.
This extension abstracts HoloLens 2 data acquisition and data transfer between processes into four components:
\begin{itemize}
    \item Source (process): continuously receives data from a single HoloLens 2 stream by calling get\_next\_packet() and sends every data frame to the interconnect.
    \item Interconnect (process): receives data frames from a single source and stores them into a ring buffer. Multiple sinks can request data from the interconnect.
    \item Sink (interface): Retrieves data frames from the interconnect. A process can manage multiple sinks, or each sink can be contained in its own process, or a combination of both.
    \item Control (process): Responsible for creating, attaching, and terminating sources, interconnects, and sinks. It is usually the main process.
\end{itemize}
The interconnect processes messages from the source, control, and sinks in a round-robin fashion.
All communications between the components are signaled by semaphores, which obviates polling mechanisms and avoids wasting CPU time busy-waiting.

The interconnect exposes the following interface to the sink objects:
\begin{itemize}
    \item get\_nearest(timestamp): returns the data frame that is closest in time to the input timestamp.
    \item get\_frame\_stamp(): returns the frame stamp of the most recent data frame.
    \item get\_most\_recent\_frame(): returns the most recent data frame.
    \item get\_buffered\_frame(frame\_stamp): returns the data frame corresponding to the input frame stamp.
\end{itemize}
The frame stamp corresponds to the global position of the data frame in the stream (0 corresponds to the first frame at the beginning of the stream).
Since the ring buffer has limited capacity, older frames will become unavailable.

The extension abstracts most of the multiprocessing details and presents the user with a simple interface.
Here is an example that acquires data from the PV and RM Depth Long-throw streams.
\begin{verbatim}
import multiprocessing as mp
import hl2ss
import hl2ss_mp
import hl2ss_utilities

# HoloLens 2 IP address
host = '192.168.1.15'

......

# A producer is a source-interconnect
# pair and the hl2ss_utilities producer
# manages the collection of producers
# for the HoloLens 2 streams
producer = hl2ss_utilities.producer()

# Initialize PV
producer.initialize_decoded_pv(
  30 * 5, # Space for 5 seconds of data
  host,
  hl2ss.StreamPort.PERSONAL_VIDEO,
  hl2ss.ChunkSize.PERSONAL_VIDEO,
  hl2ss.StreamMode.MODE_0,
  640,
  360,
  30,
  hl2ss.VideoProfile.H265_MAIN,
  1*1024*1024,
  'rgb24')

# Initialize RM Depth Long-throw
producer.initialize_decoded_rm_depth(
  5 * 5, # Space for 5 seconds of data
  host,
  hl2ss.StreamPort.RM_DEPTH_LONGTHROW,
  hl2ss.ChunkSize.RM_DEPTH_LONGTHROW,
  hl2ss.StreamMode.MODE_1)

# The sources start acquiring data and
# the interconnects add it to their
# ring buffers
producer.start()

# We use this manager to create a
# semaphore that signals when new
# data has been received
manager = mp.Manager()

# This consumer object manages a
# collection of sinks
consumer = hl2ss_utilities.consumer()

# Create a sink to read from the PV
# producer, but don't create a
# semaphore
sink_pv = consumer.create_sink(
  producer,
  hl2ss.StreamPort.PERSONAL_VIDEO,
  manager,
  None) # No semaphore

# Create a sink to read from the depth
# producer and create a semaphore
# because we want to know when a new
# depth frame arrives
sink_depth = consumer.create_sink(
  producer,
  hl2ss.StreamPort.RM_DEPTH_LONGTHROW,
  manager,
  ...) # Create new semaphore

# Semaphores can be shared by sinks of
# the same consumer, but then the user
# is responsible for checking which of
# the producers received new data

# Get the frame stamp of when the sink
# was attached to the producer. Used to
# synchronize producer and sink 1-to-1
# (not shown in this example). Must be
# called before any other sink method
sink_pv.get_attach_response()
sink_depth.get_attach_response()

while (enable):
  # Wait for new data
  sink_depth.acquire()

  # Get depth data and check that the
  # pose is valid
  data_depth = 
    sink_depth.get_most_recent_frame()
  if (not data_depth.is_valid_pose()):
    continue

  # Get the closest RGB frame
  _, data_pv =
    sink_pv.get_nearest(
      data_depth.timestamp)
  if (data_pv is None):
    continue

  # At this point we have an RGB-depth
  # image pair with pose we can use for
  # 3D reconstruction

......

# Detach sinks from producer
sink_pv.detach()
sink_depth.detach()

# Terminate producer
producer.stop()
\end{verbatim}

We use a variation of this example to perform real time TSDF integration on a client machine using Open3D \cite{pyc_open3d}.

\subsection{Interoperability}

All decoded data is returned as NumPy arrays that can be used as is with other libraries:
\begin{itemize}
    \item RM VLC: 480x640 of u8.
    \item RM Depth: 288x320 of u16 for both depth and AB.
    \item PV: height x width x 3 of u8.
    \item Microphone: 2x1024 of float.
\end{itemize}

\begin{figure}[t]
    \centering
    
    \begin{minipage}{0.49\linewidth}
        \centering
        \includegraphics[width=\linewidth]{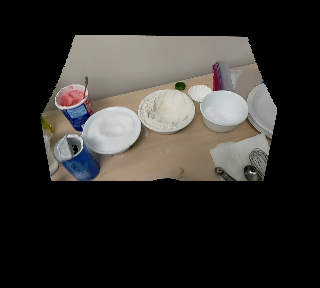} 
    \end{minipage}
    \begin{minipage}{0.49\linewidth}
        \centering
        \includegraphics[width=\linewidth]{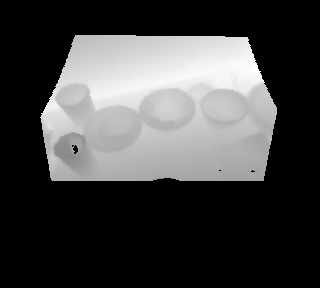} 
    \end{minipage}

    \begin{minipage}{0.49\linewidth}
        \centering
        \includegraphics[width=\linewidth]{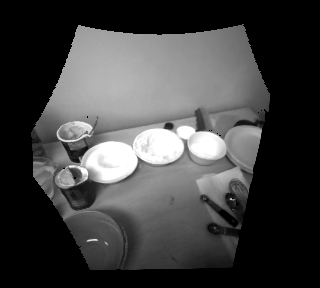} 
    \end{minipage}
    \begin{minipage}{0.49\linewidth}
        \centering
        \includegraphics[width=\linewidth]{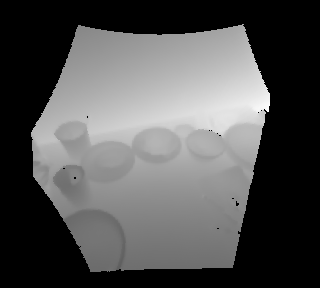} 
    \end{minipage}

   \caption{RGBD image alignment. Color can be from either the RM VLC (top) or the PV (bottom) stream.}
   \label{fig:rgbd}
\end{figure}

Additionally, our library provides utility functions for data preprocessing and storage:
\begin{itemize}
    \item Image undistort for the RM VLC and RM Depth streams (PV images have no distortion).
    \item RGBD image alignment: RM VLC + Depth or PV + Depth (AB + Depth are always aligned). See \autoref{fig:rgbd}.
    \item Associate data from different streams.
    \item Transforms between reference frames.
    \item Write/Read timestamps, data, and poses to file and save encoded audio / video to MP4.
\end{itemize}
Undistorting and aligning is performed using OpenCV \cite{pyc_opencv} and writing to MP4 uses PyAV \cite{pyc_pyav}. 
Since decoding, undistorting, and aligning are performed in real time, our framework can be integrated into real time, online systems.

\begin{figure}[t]
   \centering
   \includegraphics[width=\linewidth]{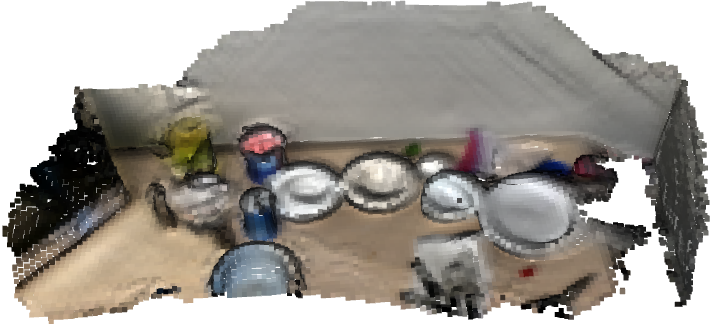}

   \caption{Point cloud generated from HoloLens 2 data using Open3D \cite{pyc_open3d}.}
   \label{fig:integration}
\end{figure}

\begin{figure}[t]
    \centering
    
    \begin{minipage}{0.49\linewidth}
        \centering
        \includegraphics[width=\linewidth]{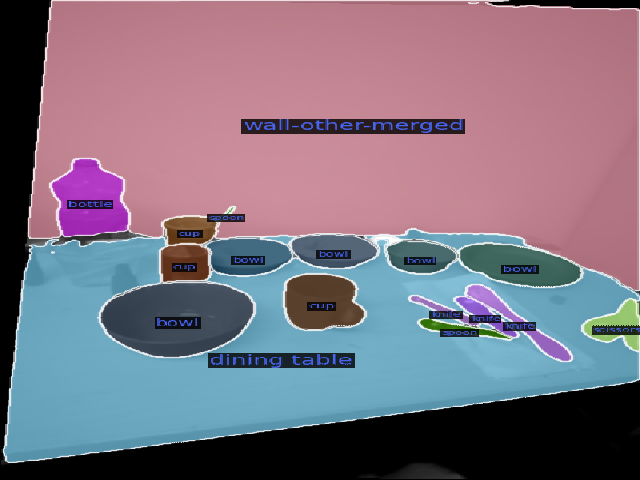} 
    \end{minipage}
    \begin{minipage}{0.49\linewidth}
        \centering
        \includegraphics[width=\linewidth]{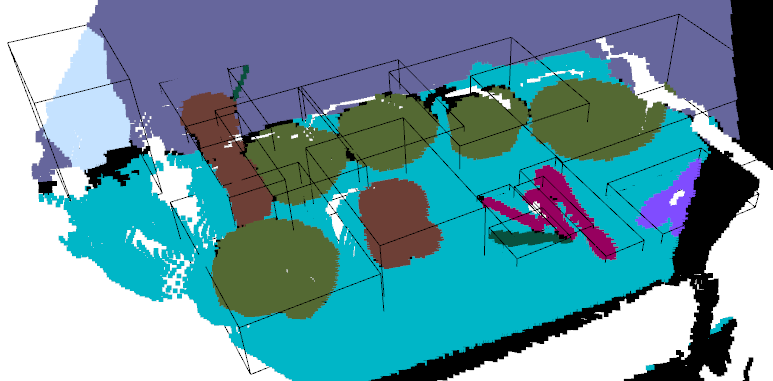} 
    \end{minipage}

   \caption{Panoptic segmentation on RM VLC data using MMDetection \cite{pyc_mmdetection} and used to 3D segment RM Depth data.}
   \label{fig:segmentation}
\end{figure}

As examples, \autoref{fig:integration} shows a point cloud created on a client machine using Open3D \cite{pyc_open3d} from HoloLens 2 rgbd-aligned data, and \autoref{fig:segmentation} shows panoptic segmentation results on an undistorted RM VLC image using MMDetection \cite{pyc_mmdetection} which is then used to segment RM Depth data.

\section{Conclusions}

We have presented a system to stream HoloLens 2 data over Wi-Fi at full framerate in real time, that enables real time online experiments on Windows, Linux, and OS X systems.
In future work, we aim to include streams for Spatial Mapping data \cite{hl2_spatial}, Scene Understanding \cite{hl2_scene}, and to give the client access to Voice Input \cite{hl2_voice}.

\section{Acknowledgments}

This work was sponsored by the Defense Advanced Research Projects Agency, the content of the information does not necessarily reflect the position or the policy of the Government, and no official endorsement should be inferred.

{\small
\bibliographystyle{ieee_fullname}
\bibliography{document}
}

\end{document}